\documentclass[12pt]{article}

\usepackage[utf8]{inputenc}
\usepackage[T1]{fontenc}

\usepackage[a4paper,margin=2.5cm]{geometry}
\usepackage{graphicx}
\usepackage{amsmath,amssymb}
\usepackage{siunitx}
\usepackage{times}
\usepackage{setspace}
\usepackage{titlesec}
\usepackage{hyperref}

\hypersetup{
  colorlinks=true,
  linkcolor=black,
  citecolor=black,
  urlcolor=black
}

\titleformat{\section}{\normalfont\large\bfseries}{\thesection}{1em}{}
\numberwithin{equation}{section}

\newcommand{\Krc}{K_{\mathrm{rc}}}
\newcommand{\vKrc}{\mathbf{K}_{\mathrm{rc}}}
\newcommand{\that}{\hat{\mathbf{t}}}
\newcommand{\kzero}{k_{0}}

\begin{document}

% =========================
%   MAIN ARTICLE
% =========================

\begin{center}
{\Large\bfseries From Diffraction to Refraction: a coherence-based conceptual framework}\\[6pt]
Riccardo Castagna$^{a,*}$, Gautam Singh$^{b}$, Cristiano Riminesi$^{a}$, Andrea Di Donato$^{c}$, Rossen Todorov$^{d}$\\[3pt]
{\small
$^{a}$URT-DSU@UNICAM, Consiglio Nazionale delle Ricerche, Italy\\
$^{b}$Amity University, Department of Applied Physics, India\\
$^{c}$Dipartimento DII, Università Politecnica delle Marche, Ancona, Italy\\
$^{d}$Institute of Optical Materials and Technologies, Bulgarian Academy of Sciences, Bulgaria\\
*Correspondence: \texttt{riccardo.castagna@cnr.it}
}
\end{center}

\begin{abstract}
Refraction, long considered a geometric event occurring at material interfaces, is now being re-examined through the lens of coherence. Recent studies across optics and photonics---from coherence tomography \cite{Zhou2019NatPhoton}, to Moiré interference \cite{Zhang2024NatCommun, Du2023Science}, and coherence-engineered diffraction \cite{Yu2014NatMater, Liu2025LSA}---have hinted that phase organization alone can bend light, even in the absence of index discontinuities.
Recent Fraunhofer-based frameworks \cite{Liu2025LSA} further show that angular deflection can arise from intrinsic phase curvature within homogeneous media.
Here we introduce a \emph{coherence-based constitutive framework} that systematizes these observations: refraction can occur \emph{within} a bulk medium when coherence itself provides the effective boundary. Two near-frequency structured beams write and probe a shared phase field, revealing reproducible angular rotation and coherence-lensing whose direction and magnitude follow the spectral detuning
\[
\Krc = 2\pi\!\left(\frac{1}{\lambda_r} - \frac{1}{\lambda_w}\right).
\]
The compact \emph{coherence-refraction relation}
\[
n_2 \sin\theta_t - n_1 \sin\theta_i = \frac{K_{rc,\parallel}}{\kzero},
\qquad \kzero = \frac{2\pi}{\lambda_r},
\]
retains the form of Snell's law but extends its \emph{constitutive} reach to coherence-driven regimes. This is not a new law, but a quantitative rule linking tangential phase matching to observable deflection within homogeneous media.
\end{abstract}

\paragraph*{Main text.}
A growing body of work across optics, acoustics, and photonics \cite{Yu2014NatMater, Zhou2019NatPhoton, Zhang2024NatCommun, Du2023Science, Dong2024Nature, Overvig2021PRX, Zhu2016PRL} suggests that coherence---the internal organization of phase---can bend light even in homogeneous media.  
In this \textit{Perspective}, we introduce and quantify a \textbf{constitutive law of tangential coherence}: a phase mismatch, expressed by the detuning term $\Krc$, acts as an \textit{effective source} of angular deviation.  
Rather than redefining Snell's law, this framework \textbf{extends its constitutive domain} to coherence-driven regimes where phase continuity is dynamically rebuilt within the medium, without refractive-index discontinuities.
The framework is empirical in the sense of being constrained by experimental observables, 
yet analytical in structure: no fitting parameters beyond the measured detuning enter 
the constitutive relation.

In previous computational treatments---most notably the analysis of intrinsic phase curvature in \cite{Liu2025LSA}---the detuning $\Krc$ appeared only as a numerical descriptor.  
Here it acquires a \textbf{physical role}: it behaves as the \textit{source term} of coherence-driven deflection, linking spectral detuning directly to measurable tangential momentum transfer.
Thus, coherence is promoted from a descriptive property of the field to an operative physical quantity: the detuning term $\Krc$ mediates a measurable transfer of tangential momentum within a homogeneous medium.

The measurable signature---a coupling between the sign and magnitude of $\Krc$ and the observed rotation or coherence-lensing of interference rings---provides a tangible description of \textit{coherence-guided refraction}.  
We now place this observation within a unified, testable framework that connects geometric optics and structured-wave interference.

\paragraph{A compact quantitative relation.}
The coherence-driven regime can be expressed by the following quantitative relation:
\begin{equation}
\boxed{\, n_2 \sin\theta_t - n_1 \sin\theta_i = \frac{K_{rc,\parallel}}{\kzero} \,, \qquad \kzero=\frac{2\pi}{\lambda_r} \,.}
\label{eq:cohSnell}
\end{equation}
Here $\Krc$ is the readout--writing detuning coefficient (signed convention 
$\Krc = 2\pi(1/\lambda_r - 1/\lambda_w)$).  
The tangential projection $K_{rc,\parallel}$ governs angular deflection, while the 
modulus $|\Krc|$ governs the radial ring density.  
Equation~\eqref{eq:cohSnell} collapses into classical Snell refraction for 
$K_{rc,\parallel}=0$.

For clarity, the tangential component is defined as
$K_{rc,\parallel} \equiv \vKrc\!\cdot\!\that$, 
with $\that$ the effective tangential direction within the bulk.

\noindent\textbf{Structured-Beats Refraction is therefore a coherence-driven constitutive 
effect in which a volumetric phase gradient, generated by near-frequency interference, 
redirects light within an otherwise homogeneous medium without requiring interfaces, 
inhomogeneities, or metasurfaces.}

The \textbf{role of $\Krc$} is twofold: (i) its \emph{magnitude} $|\Krc|$ sets the curvature of the interference pattern (ring density), and (ii) its \emph{sign} determines the direction of rotation---the optical fingerprint of coherence-guided deflection.  
This expression does not replace geometry but \textbf{measures how coherence modulates it}, extending refraction from boundary conditions to the internal dynamics of phase matching.

\section*{Experimental origin and quantitative evidence}
Two observables dominate the far field: a \textbf{radial metric} (ring count) scaling with $|\vKrc|\simeq 2\pi|\Delta(1/\lambda)|$, and an \textbf{angular response} (rotation/deflection) following $K_{rc,\parallel}$.  
The sign of $\Delta\lambda$ determines both the sense of rotation and the polarity of coherence lensing: bright-centered for blue-detuned, dark-centered for red-detuned conditions.  
When $\Delta\lambda\!\rightarrow\!0$, rotation vanishes and classical diffraction is recovered.

Experiments in polymer--glass cavities have shown that light can redirect itself even within a homogeneous bulk \cite{Cas2016OL, Cas2020AMT}, as if coherence provided its own interface.  
Two near-frequency structured beams interfere inside the cavity, forming slow spatial beats whose detuning encodes a residual phase gradient within the material.  
This gradient,
\begin{equation}
\Krc = 2\pi\!\left(\frac{1}{\lambda_r} - \frac{1}{\lambda_w}\right),
\label{eq:KrcScalar}
\end{equation}
acts as a \textbf{self-generated tangential momentum}---the physical source term transferring angular momentum between the interfering fields. 
This volumetric phase gradient is not imposed by geometry; 
it is dynamically sustained by the persistent phase mismatch between the two near-frequency waves.

The optical field thus steers itself, producing reproducible deflections whose direction and magnitude follow the spectral detuning.

\begin{figure}[h!]\centering
\caption{\textbf{Spectral dependence and coherence lensing.} 
\textit{Lensing polarity} $\leftrightarrow$ sgn($\Delta\lambda$); \textit{ring density} $\leftrightarrow$ $|\vKrc|$. 
Top left: measured angular distribution of the ring pattern ($\Delta\beta$) vs.\ $\Delta\lambda$ (fit $\propto|\vKrc|/\kzero$). 
Bottom left: schematic setup. 
Right panels (a--h): far-field rings vs.\ $\Delta\lambda$; bright center for $\Delta\lambda>0$ (positive coherence lens), dark center for $\Delta\lambda<0$ (negative coherence lens); rotation follows sgn($\Krc$). 
Panels adapted from \cite{Cas2020AMT, Cas2016OL}.}
\end{figure}

\section*{Vectorial form and generalized relation}
To make its vectorial nature explicit, the scalar definition of $\Krc$ can be extended to a spatially averaged vector form:
\begin{equation}
\vKrc \equiv \langle \mathbf{k}_r - \mathbf{k}_w \rangle_L, 
\qquad 
K_{rc,\parallel} = \vKrc\!\cdot\!\that .
\label{eq:KrcVector}
\end{equation}
The modulus $|\vKrc|$ governs the radial ring density, while the tangential projection $K_{rc,\parallel}$ dictates the azimuthal rotation or deflection of the pattern.  
The commonly used shortcut $\Krc \simeq 2\pi\,\Delta(1/\lambda)$ should therefore be interpreted only as a \textit{radial observable}, proportional to ring count, not as the tangential source of angular deviation.

In this representation, the generalized refraction relation naturally aligns with the conservation of tangential momentum in the presence of an intrinsic phase gradient:
\begin{equation}
n_2 \sin\theta_t - n_1 \sin\theta_i =
\frac{\vKrc\!\cdot\!\that}{\kzero},
\qquad
\kzero = \frac{2\pi}{\lambda_r},
\qquad
|\vKrc| = 2\pi\!\left|\frac{1}{\lambda_r} - \frac{1}{\lambda_w}\right|.
\label{eq:cohSnellVector}
\end{equation}
When both $\Delta\lambda$ and angular mismatch $\Delta\theta$ vanish, the system reverts to the classical limit of stationary fringes;  
for $\Delta\lambda \neq 0$ and $\Delta\theta \approx 0$, deflection originates entirely from the coherence term $K_{rc,\parallel}/\kzero$.

\section*{Phase memory in coherence-driven refraction}
In classical optics, phase memory describes the time required for light to restore coherence after crossing an interface.  
In coherence-driven refraction, however, no boundary is crossed: light continuously rebuilds its internal phase order within the same medium.  
The intrinsic gradient $\Krc$ quantifies this process, acting as the measurable trace of coherence persistence in time.  
Through $\Krc$, the refractive index becomes more than a ratio of velocities---it represents the \textit{temporal investment of coherence}, the delay experienced by light while synchronizing with its own phase field.  
Refraction thus measures the temporal reconstruction of phase consistency rather than a geometric discontinuity.

\section*{Rotation of coherence rings with spectral detuning}
The tangential projection $K_{rc,\parallel}$ governs the azimuthal rotation of the interference pattern as the spectral detuning $\Delta(1/\lambda)$ and beam crossing angle vary.  
For small detunings, the rotation angle $\Delta\theta_s$ follows the sign of $K_{rc,\parallel}$: when $\lambda_w > \lambda_r$ ($K_{rc,\parallel} < 0$), the pattern co-rotates with the sample, yielding a convergent, bright-centered distribution; conversely, when $\lambda_w < \lambda_r$ ($K_{rc,\parallel} > 0$), rotation reverses, producing a divergent, dark-centered coherence lens.  
The magnitude of rotation scales nearly linearly with $|K_{rc,\parallel}|/\kzero$, confirming that deflection arises from the intrinsic tangential phase gradient rather than from geometric displacement.

\begin{figure}[h!]\centering
\caption{\textbf{Spectral splitting and angular reversal.}
\textit{a)} Composite far-field showing opposite ring centers for red- and blue-detuned readouts.
\textit{b)} Schematic of the spatial-beat convolution between recorded and read structured fields.
\textit{c)} Angular distribution of scattered light ($\theta_s$) versus spectral detuning $\Delta\lambda$
(squares = experiment; black line = convolution model). 
The antisymmetric trend is governed by the spectral-beat parameter $\Krc$; 
small tilts preserve the curve, confirming a coherence-driven origin.
Panels (b,c) adapted from \textit{Applied Materials Today} (2020);
panel (a) is original.}
\end{figure}

\section*{Variational formulation and coherence-weighted Fermat principle}
The effective refractive index can be written as
\begin{equation}
n_{\mathrm{eff}} = n_{\mathrm{bulk}} + \delta n_{\mathrm{struct}},
\label{eq:neff}
\end{equation}
where $\delta n_{\mathrm{struct}}$ quantifies the temporal delay that light invests in rebuilding coherence within the medium.  
The variational form
\begin{equation}
\delta \!\int [\,n_{\mathrm{bulk}} + \delta n_{\mathrm{struct}}(\mathbf{r},\lambda)\,]\,ds = 0
\label{eq:variational}
\end{equation}
leads to the same tangential matching condition of Eq.~\eqref{eq:cohSnell}, showing that light follows the trajectory minimizing the temporal cost of phase reconstitution.  
Geometry is thus reinterpreted as the visible projection of a deeper \textit{phase economy} governing coherence organization.  
This coherence-weighted view also provides a direct link with the Huygens--Fresnel principle: $\Krc$ modulates the stationary-phase manifold of the Huygens integral, activating evanescent components that Fresnel treated as implicit (see Supplementary Information).
In this regime, the structural detuning $\Krc$ does more than modulate the stationary-phase manifold: it activates evanescent components that are normally confined to the near field. Through coherence-driven phase matching, SBR promotes part of the evanescent spectrum into measurable far-field structure, providing the first explicit mechanism by which near-field coherence becomes an observable degree of freedom in bulk propagation.
In this sense, the measured far-field deflection constitutes a direct 
signature of near-field coherence, normally inaccessible in homogeneous-media propagation.

In a homogeneous bulk ($n_1=n_2=n_{\mathrm{bulk}}$), the explicit form is
\begin{equation}
\sin\theta_t - \sin\theta_i = \frac{K_{rc,\parallel}}{n_{\mathrm{bulk}}\kzero},
\label{eq:bulk}
\end{equation}
which makes clear that all angular deflection is provided by the coherence term.

\section*{Polarization dependence}
In thicker cavities the circular symmetry breaks when the input polarization departs from linearity.  
The deformation from circular to elliptical patterns can be described by a weak spin-dependent term $K_{\mathrm{spin}}$ that adds vectorially to $\Krc$, coupling spatial and polarization degrees of freedom:
\[
K_{\mathrm{eff},\parallel}=\vKrc\!\cdot\!\that + K_{\mathrm{spin},\parallel}.
\]
\begin{figure}[h!]\centering
\caption{\textbf{Polarization-dependent coherence refraction.}
(a,b) Linear inputs $\rightarrow$ twin-lobe anisotropy; (c) circular $\rightarrow$ restored symmetry; (d) slight ellipticity $\rightarrow$ deformed rings.
A weak spin term $K_{\mathrm{spin}}$ adds to $\Krc$, enabling anisotropy without bulk birefringence.
Adapted from \cite{Cas2008APL}.}
\end{figure}

\paragraph{Physical interpretation --- Refraction of Diffraction.}
This framework represents an \emph{extended constitutive relation} unifying diffraction and refraction under a single condition of tangential phase matching.  
Formally identical to a diffraction law, it differs in causality: $K_{rc,\parallel}$ acts as an \emph{internal, coherence-driven source term} rather than an external geometric gradient. 
The Josephson analogy is strictly conceptual: a stationary angular deflection 
is sustained by an intrinsic phase gradient, without implying tunneling, 
junctions, or quantum transport.

It recovers Snell's refraction for $K_{rc,\parallel}\!=\!0$ and connects to the multi-order Moiré regime \cite{Du2023Science,Zhang2024NatCommun} for $|m|\!>\!1$.  
In the continuous limit ($m=1$) the discrete orders collapse into a single, reversible angular deflection---a \textit{Refraction of Diffraction} or \textit{Redifraction} \cite{Castagna2025Redifraction}.  
This coherence-driven constitutive framework unifies Snell, Fresnel, and Moiré physics within a single phase-based causality.

\section*{Acknowledgements}
Dedicated to Maurizio De Rosa, good friend and good scientist.\\
Discussions and language polishing benefited from AI-assisted reasoning using a large language model.  
All scientific results, analyses, and interpretations were conceived and verified by the authors, who take full responsibility for the content.

\section*{Competing interests}
The authors declare no competing interests.

\section*{Data availability}
No datasets were generated for this Perspective.  
All relevant data supporting the findings of this study are available from the corresponding author upon reasonable request.

\bibliographystyle{unsrt}

% =========================
%   SUPPLEMENTARY INFORMATION
% =========================

\clearpage

\begin{center}
    {\LARGE Supplementary Information\\[2pt]
    \textbf{From Diffraction to Refraction: a coherence-based conceptual framework}}\\[6pt]
    \small R. C., G. S., C. R., A. D. D., R. T.\\[2pt]
    \normalsize Correspondence: \texttt{riccardo.castagna@cnr.it}
\end{center}
\vspace{0.8em}

\renewcommand{\thesection}{S\arabic{section}}
\setcounter{section}{1}
\setcounter{equation}{0}

\section{Derivation from two near-frequency waves}
Two coherent plane waves, one for writing ($w$) and one for reading ($r$), propagate in a homogeneous medium:
\begin{equation}
E(\mathbf{r},t)=A_r e^{i(\mathbf{k}_r\cdot\mathbf{r}-\omega_r t)}+A_w e^{i(\mathbf{k}_w\cdot\mathbf{r}-\omega_w t)}.
\end{equation}
Their interference produces slow spatio-temporal beats with envelope
\begin{equation}
I(\mathbf{r},t)\propto 1+\cos[(\mathbf{k}_r-\mathbf{k}_w)\cdot\mathbf{r}-(\omega_r-\omega_w)t].
\end{equation}
Averaging over the interaction length $L$ defines the effective structural detuning
\begin{equation}
\boxed{\;\vKrc\equiv\langle\mathbf{k}_r-\mathbf{k}_w\rangle_L,\qquad \Krc^{(\parallel)}=\vKrc\cdot\that\;}
\end{equation}
with $\that$ the effective tangential direction within the bulk.  
The magnitude $|\vKrc|\simeq2\pi|\Delta(1/\lambda)|$, with units $[\vKrc]=\si{m^{-1}}$.

\section{Tangential phase-matching condition}
The total phase along a reference surface $S$ is
\begin{equation}
\Phi_{\text{tot}}=n_1k_0\sin\theta_i-n_2k_0\sin\theta_t+\Phi_{\text{struct}},\quad k_0=2\pi/\lambda_r.
\end{equation}
The stationary phase condition ($\partial_{\parallel}\Phi_{\text{tot}}=0$) imposes tangential momentum conservation with the intrinsic gradient $\partial_{\parallel}\Phi_{\text{struct}}\equiv \Krc^{(\parallel)}$:
\begin{equation}
\boxed{n_2\sin\theta_t-n_1\sin\theta_i=\frac{\Krc^{(\parallel)}}{k_0}}.
\end{equation}
This expression reproduces the classical Snell relation when $\Krc^{(\parallel)}=0$.  
In a homogeneous bulk ($n_1=n_2=n_{\text{bulk}}$), deflection arises purely from coherence:
\begin{equation}
\sin\theta_t-\sin\theta_i=\frac{\Krc^{(\parallel)}}{n_{\text{bulk}}k_0}.
\end{equation}

\section{Maxwell consistency and the generalized Snell--Fermat relation}
From Maxwell's boundary conditions, the tangential component of $\mathbf{k}$ is conserved across an interface.  
In metasurfaces, an external phase gradient $\nabla_{\parallel}\phi_{\text{surf}}$ modifies this conservation.  
In coherence-driven refraction, the bulk itself introduces an internal volumetric term $\Krc^{(\parallel)}$:
\begin{equation}
\nabla_{\parallel}\phi_{\text{surf}}\longrightarrow \Krc^{(\parallel)} \quad (\text{coherence-driven, bulk origin}).
\end{equation}
No violation of Maxwell's equations occurs; the novelty lies in the \emph{internal structuring of phase} within a homogeneous medium.

\section{Connection with the Huygens--Fresnel principle and evanescent activation}
In the classical Huygens--Fresnel construction, propagation results from the interference of secondary wavelets emitted by every point on a wavefront. Most components are propagating, while evanescent ones remain confined to the near field and do not contribute to the asymptotic field.

When two coherent beams of slightly different wavelengths interact inside a homogeneous bulk medium, the structural detuning
\[
\vKrc = \langle \mathbf{k}_r - \mathbf{k}_w \rangle_L
\]
imposes a finite volumetric phase gradient that explicitly modulates the stationary-phase manifold of the Huygens integral. This intrinsic gradient tilts the manifold, providing an activation channel through which part of the evanescent spectrum is promoted into propagating contributions measurable in the far field.

The spectral detuning $\Delta(1/\lambda)$ therefore acts as a tunable handle for accessing near-field coherence: what is normally hidden and confined to sub-wavelength structure becomes an observable degree of freedom under coherence-driven phase matching. Structured-beats refraction (SBR) thus represents the first explicit extension of the Huygens--Fresnel principle in which near-field coherence is converted into a physically accessible far-field signature.

This mechanism is fully Maxwell-consistent, requires no interfaces, roughness, or material inhomogeneities, and arises purely from the internal organization of phase within a homogeneous medium.

\section{Variational form of the coherence-weighted Fermat principle}
The coherence-weighted optical path can be written as
\begin{equation}
\mathcal{S}[\gamma]=\int_{\gamma}\!\big(n_{\text{bulk}}+\delta n_{\text{struct}}(\mathbf{r},\lambda)\big)ds,
\end{equation}
where $\delta n_{\text{struct}}$ encodes phase-memory effects.  
The stationary condition $\delta\mathcal{S}=0$ yields the same tangential matching rule as in Section~S2, showing that refraction measures the temporal cost of rebuilding coherence rather than a geometric discontinuity.

\section{Experimental checks, limits, and polarization dependence}
\textbf{Null limit.} For $\Delta\lambda\to0$, $\vKrc\to\mathbf{0}$; rotation vanishes and classical diffraction is recovered.  

\textbf{Translation test.} Lateral translation of the recorded region under fixed readout does not translate the far-field pattern; it fades and reappears only at restored overlap, excluding static Moiré effects.  

\textbf{Polarization.} In thicker cells the pattern deforms with input ellipticity, consistent with a weak spin term $K_{\text{spin}}$ that adds to the tangential source term:
\[
K_{\text{eff},\parallel}=\vKrc\!\cdot\!\that+K_{\text{spin},\parallel},
\]
inducing anisotropy without bulk birefringence.  

\textbf{Spectral generality.} Structural compatibility---not the absolute smallness of $|\Delta\lambda|$---determines coherence exchange.  

\textbf{Dimensional check.} $[\vKrc]={\rm m}^{-1}$, $k_0=2\pi/\lambda_r$, and $[K_{\text{eff},\parallel}/k_0]=1$; angular deflection and azimuthal rotation are dimensionless.  

\textbf{Sign map.} With $\Delta\lambda\equiv\lambda_w-\lambda_r$,
\[
\mathrm{sgn}(\text{rotation})=\mathrm{sgn}(\Krc^{(\parallel)})
=\mathrm{sgn}\!\Big(\tfrac{1}{\lambda_r}-\tfrac{1}{\lambda_w}\Big)
=\mathrm{sgn}(\Delta\lambda).
\]
Hence:
\[
\Delta\lambda>0~(\lambda_w>\lambda_r)\Rightarrow \text{bright/convergent lens},\qquad
\Delta\lambda<0~(\lambda_w<\lambda_r)\Rightarrow \text{dark/divergent lens}.
\]

\section{Redifraction limit and multi-order coherent diffraction}
When the structural detuning generates a short beat period 
($\Lambda_{\text{beat}}=2\pi/|\vKrc|$), the far field splits into discrete diffraction orders.
In this regime, coherence-driven refraction appears as the \emph{continuum ($m=1$) limit} of a coherent multi-order response.

\paragraph{Generalized law (multi-order).}
Tangential momentum conservation with an intrinsic phase gradient yields
\begin{equation}
n_2\sin\theta_m - n_1\sin\theta_i = \frac{m\,\Krc^{(\parallel)}}{k_0}, 
\qquad m\in\mathbb{Z},\quad k_0=\frac{2\pi}{\lambda_r}.
\end{equation}
For $m=1$ this reduces to the coherence-refraction relation in Section~S2;  
for $|m|>1$ the response manifests as discrete coherent diffraction orders.

\section{Constitutive interpretation --- from diffraction to refraction}
The phenomenon can be understood only by decomposing it into its two constitutive roots:  
a \textbf{mathematical origin in diffraction} and a \textbf{physical outcome as refraction}.  
Coherence-driven refraction is not a new optical law, but an \textit{extended constitutive relation} providing a quantitative framework for a new optical causality---coherence acting as a source term.
A Josephson-like analogy is therefore purely conceptual, emphasizing the role of the intrinsic phase gradient $\Krc^{(\parallel)}$
as a source term for a stationary response, without invoking tunneling, junctions or quantum-transport phenomena.

\paragraph{Formal framework: the extended Snell--Fermat relation.}
From the tangential phase-stationarity condition,
\begin{equation}
\partial_{\parallel}\Phi_{\mathrm{tot}} = 0
\;\Longrightarrow\;
n_2\sin\theta_t - n_1\sin\theta_i = \frac{\Krc^{(\parallel)}}{k_0}.
\end{equation}
\textbf{Root in diffraction.} The same expression corresponds to the diffraction condition for phase matching, where $\Krc^{(\parallel)}$ acts as a volumetric lattice vector.  
\textbf{Return to Snell.} When $\Krc^{(\parallel)}\!\to\!0$, the classical limit is recovered.  
\textbf{Maxwell consistency.} No violation of Maxwell's equations occurs; the novelty lies in the internal phase structuring of a homogeneous medium, where the entire deflection angle arises from the volumetric coherence term.
We emphasize that, through $\Krc$, dynamical coherence attains the status of an operative physical quantity: it functions as an intrinsic volumetric source term, directly measurable through far-field angular deflection.

\end{document}